\documentclass[a4paper,11pt,fleqn]{article}
\usepackage{latexsym}
\usepackage{amsmath} \usepackage{amsthm}

\begin{document}
\newcommand{\itt}{\textit}
\def\ctg{\qopname\relax o{ctg}} \newcommand{\scrL}{\mathcal{L}}
\newcommand{\scrT}{\mathcal{T}} \newcommand{\scrD}{\mathcal{D}}
\newcommand{\scrE}{\mathcal{E}} \newcommand{\scrJ}{\mathcal{J}}
\newcommand{\scrX}{\mathcal{X}} \newcommand{\scrQ}{\mathcal{Q}}
\newcommand{\scrS}{\mathcal{S}} \newcommand{\scrR}{\mathcal{R}}
\newcommand{\scrK}{\mathcal{K}} \newcommand{\scrO}{\mathcal{O}}
\newtheorem{definition}{Definition} \newtheorem{lemma}{Lemma}
\numberwithin{equation}{section}

\title{\bf The 3+1 decomposition of Conformal Yano-Killing tensors and
``momentary'' charges for spin-2 field}
\author{Jacek Jezierski
and Szymon Migacz\\ Department of Mathematical Methods in
Physics, \\ University of Warsaw, ul. Ho\.za 74, 00-682 Warsaw,
Poland }

\maketitle
\begin{abstract}
The ``fully charged'' spin-2 field
solution is presented. This is an analog of the Coulomb solution
in electrodynamics and represents the ``non-waving'' part of the
spin-2 field theory.

Basic facts and definitions of the
spin--2 field and conformal Yano-Killing tensors are introduced.
Application of those two objects provides a precise definition
of quasi-local gravitational charge. 
Next, the  3+1 decomposition leads to the
construction  of the momentary
gravitational charges on initial surface which is applicable for Schwarzschild-like spacetimes.
\end{abstract}

\section{Introduction}
\paragraph{Purpose of this paper}
The
charges associated with the gravitational field play a
significant role in General Theory of Relativity. Our goal is to
provide a definition of energy, momentum and angular momentum of
a gravitational field using analogies between linearized
gravitation and classical Maxwell's electrodynamics. 
Following \cite{GRG27} we shortly remind how, using
the geometrical objects characterizing a gravitational field
(spin-2 field $W_{\alpha \beta \mu \nu}$ and conformal
Yano-Killing tensors), one can obtain the quasi-local charges in
Minkowski spacetime. 

Using a 3+1 decomposition of the spacetime, we propose to define the momentary charges in terms of
natural initial value tensors: the electric and the magnetic part of
the spin-2 field and the initial value data for CYK tensor -- Conformal Killing Vector field (CKV).
This is applicable for the Schwarzschild spacetime which possesses much less CYK
tensors than
Minkowski (less ``hidden'' symmetries).
Moreover, this construction works for any initial data which is
conformally flat, in particular, for any spherically symmetric
three-metric. Obviously, it leads to quasi-local quantities defined on a Cauchy surface.

\paragraph{Remarks about the notation}
In this paper we assume
that the metric $g$ has a positive signature $(-,+,+,+)$ and we are
using units, where $c=G=1$. Antisymmetrization of a tensor is
denoted with square brackets, symmetrization with round
brackets. A four-dimensional covariant derivative (for a
Levi-Civita connection) is denoted by $\nabla$ or with a
semicolon. A three-dimensional covariant derivative is on the
other hand denoted by $\overset{3}{\nabla}$ or with a vertical
segment. Greek letters $\alpha$,$\beta$,...,$\mu$,$\nu$,... are
indices with values in the set $\{0,1,2,3\}$, whereas Latin
letters $i$,$j$,... assume values in $\{1,2,3\}$.

\subsection{Spin-2 field} We will begin with defining a spin-2
field, which can also be identified with a Weyl tensor in
linearized gravitation. \begin{definition} Tensor field
$W_{\alpha \beta \mu \nu}$ is called a spin-2 field if and only
if the following conditions are fulfilled:
\begin{align}
\textrm{algebraic:} & \begin{subequations} \begin{cases}
W_{\alpha \beta \mu \nu}=W_{\mu \nu \alpha \beta}=W_{[\alpha
\beta] [\mu \nu]},\\ W_{\alpha [\beta \mu \nu]}=0, \\ g^{\alpha
\mu} W_{\alpha \beta \mu \nu}=0,\\ \end{cases}
\end{subequations}\\ \textrm{differential:} & \quad \nabla_{[
\lambda}W_{\alpha \beta ] \mu \nu} =0. \label{rozniczkowy}
\end{align} \end{definition} 
$W_{\alpha \beta \mu \nu}$ is antisymmetric in the first and in
the second pair of indices, so we can define two dual tensors,
for the first and the second pair of indices respectively:
\begin{equation} ^*W_{\rho \sigma \mu
\nu}:=\frac{1}{2}\epsilon_{\rho \sigma}{}^{\alpha \beta}
W_{\alpha \beta \mu \nu}, \qquad \qquad W^*{}_{\rho \sigma \mu
\nu}:=\frac{1}{2}W_{\rho \sigma \alpha \beta}\epsilon^{\alpha
\beta}{}_{\mu \nu}. \label{Wdualne} \end{equation} A contraction of
a spin-2 field with a normed vector $n^\mu$ perpendicular to the
foliation $\Sigma_t$ allows us to define two new tensors: the
electric and the magnetic part of Weyl tensor.\\
Electric
part:
\begin{equation}
	E_{\alpha\beta}:= W_{\alpha\mu\nu\beta}n^\mu n^\nu\,,
\end{equation}
and magnetic part:
\begin{equation}
	H_{\alpha\beta}:= W^*{}_{\alpha\mu\nu\beta}n^\mu n^\nu\,.
\end{equation}
Both defined tensors are symmetric (because of
the symmetry of the  field $W$) and traceless. Applying a
contraction with $n^\mu$ to the electric or the magnetic part
gives us zero (because of the antisymmetry of $W$ in pairs of
indices): \begin{equation} E_{\mu \nu} n^\mu = H_{\mu \nu} n^\mu
=0 \, .\end{equation} The property (\ref{rozniczkowy}) can be
rewritten in an equivalent form (\cite{GRG27}): \begin{eqnarray}
\label{rpola} \nabla_{[ \lambda}W_{\alpha \beta ] \mu \nu} =0
\iff \nabla^\alpha W_{\alpha \beta \mu \nu} =0 \iff \label{11}
\\ \iff \nabla_{[ \lambda}{}^*W_{\alpha \beta ] \mu \nu} =0 \iff
\nabla^\alpha {}^*W_{\alpha \beta \mu \nu} =0. \end{eqnarray}
\subsection{Conformal Yano-Killing tensors} Previously
introduced field $W$  is a tensor field with four indices. If we
would like to follow the analogy with the classical
electrodynamics and the Maxwell tensor, we need to define a
tensor field that has two indices. Additionally, we would be
pleased if the integral of this new object didn't depend on
the choice of the two-dimensional integration surface. We will
show that the contraction of a field $W$  with a conformal
Yano-Killing tensor has both mentioned properties. \vskip 2mm
\begin{definition} Antisymmetric tensor $Q_{\mu \nu}$ is a
conformal Yano-Killing tensor (CYK tensor) for a metric $g_{\mu
\nu}$ iff:
\begin{equation} \scrQ_{\mu \nu \kappa}(Q,g)=0,
\end{equation} \end{definition}
\noindent \textit{where}  
$\displaystyle \scrQ_{\lambda \kappa \sigma}(Q,g):=Q_{\lambda
\kappa ; \sigma}+Q_{\sigma \kappa ; \lambda}-
\frac{2}{n-1}(g_{\sigma \lambda} {Q^\nu}_{\kappa ;\nu}+
g_{\kappa ( \lambda} Q_{\sigma )}{}^\mu{}_{;\mu})$.\\[1ex] 
We are only considering four-dimensional case ($n=4$),
therefore: \begin{equation} \scrQ_{\lambda \kappa
\sigma}(Q,g):=Q_{\lambda \kappa ; \sigma}+Q_{\sigma \kappa ;
\lambda}- \frac{2}{3}(g_{\sigma \lambda} {Q^\nu}_{\kappa ;\nu}+
g_{\kappa ( \lambda} Q_{\sigma )}{}^\mu{}_{;\mu}). \end{equation}
\section{Minkowski spacetime}
\paragraph{Introduction} The flat Minkowski spacetime is the
simplest possible background for linearized gravitation.
Reasoning presented below will lead us to exact
results which are very important. It will also help us gain some insight which will be
useful considering less trivial metrics. Accordingly to the
conditions imposed in the introduction, we will be looking for a
natural object which can be integrated over two-dimensional surfaces and will
fulfill Gauss Law. 
\paragraph{Quasi-local charges -- definition}
Now we combine a spin-2 field and conformal
Yano-Killing tensor to 
define
gravitational charge in the Minkowski spacetime.\\ 
\noindent Let us assume that $W_{\alpha \beta \mu \nu}$ is a
spin-2 field and $Q_{\mu \nu}$ is any antisymmetric tensor. 
The new tensor field\footnote{Intentionally we are using the same
symbol as for the Maxwell tensor.} $F_{\mu \nu}$:
\begin{equation} F_{\mu \nu}(W,Q):=W_{\mu \nu \alpha
\beta}Q^{\alpha \beta} \end{equation} 
obeys the following
\begin{lemma}Divergence of a  tensor $F_{\mu \nu}$ takes the form: \label{lemat1}
\[ \nabla_\nu F^{\mu \nu}(W,Q)=\frac{2}{3}W^{\mu \nu \alpha
\beta}\scrQ_{\alpha \beta \nu}. \] \end{lemma} \vskip 3mm
\noindent If $Q_{\mu \nu}$ is a CYK tensor, then $\scrQ_{\alpha
\beta \nu}=0$ and Lemma \ref{lemat1} gives $\nabla_\nu
F^{\mu \nu}(W,Q)=0$.

\noindent
Let $V$ be a three-dimensional volume with a boundary $\partial V$, therefore:
\begin{equation} \int_{\partial V} F^{\mu \nu} (W,Q) d
\sigma_{\mu \nu}= \int_V \nabla_\nu F^{\mu \nu} (W,Q) d
\Sigma_\mu =0. \label{diwergencja} \end{equation} \vskip 3mm
\noindent Using this last equality we can claim that each CYK
tensor $Q_{\mu \nu}$ defines a charge connected with a spin-2
field. That's because the flux of a tensor $F^{\mu \nu}$ through
any two closed two-dimensional surfaces $S_1$ and $S_2$ is
equal if there exists a three-dimensional volume $V$ with a
boundary $\partial V$ equal to the sum of the surfaces $S_1$ and
$S_2$. 
\paragraph{CYK tensors in Minkowski spacetime}
The basis of the space of solutions for the
equation $\scrQ_{\alpha \beta \mu}=0$ (i.e. the basis of CYK
tensors) in Minkowski spacetime consists of the following 20 tensors
(cf. \cite{GRG27}):
\begin{equation} \scrT_0 \wedge \scrT_k, \hskip 3mm \scrT_0
\wedge \scrD, \hskip 3mm \scrT_k \wedge \scrD, \hskip 3mm \scrD
\wedge \scrL_{0 k} -\frac{1}{2} \eta(\scrD,\scrD)\scrT_0 \wedge
\scrT_k, \label{baza} \end{equation}
\[ *(\scrT_0 \wedge \scrT_k), \hskip 3mm *(\scrT_0
\wedge \scrD), \hskip 3mm *(\scrT_k \wedge \scrD), \hskip 3mm
*(\scrD \wedge \scrL_{0 k} -\frac{1}{2} \eta(\scrD,\scrD)\scrT_0
\wedge \scrT_k)\, , \] where $\scrD:=x^\mu \partial_\mu$,
$\scrT_\mu:=\partial_\mu$, $\scrL_{\mu \nu}:=x_\mu \partial_\nu
- x_\nu \partial_\mu$. 
 Each CYK tensor
in Minkowski spacetime can be expressed as a linear combination
(with constant coefficients) of these twenty tensors.
\paragraph{Gravitational charges in a 3+1 decomposition} \label{lad}
We begin with a following simple
observation. \begin{lemma}{} \label{lemat2} Each CYK tensor in
Minkowski spacetime can be expressed in a following way:
\begin{equation} Q=a(t) \scrT_0 \wedge X + b(t) *(\scrT_0 \wedge
Y), \label{obserwacja} \end{equation} where $X$, $Y$ are (three-dimensional)
conformal Killing fields; $a(t)$, $b(t)$
are functions of time only. \end{lemma} \noindent We will prove
this Lemma by giving the proper decomposition of the basis
tensors described by formulae (\ref{baza}).

\noindent
Let us introduce the basis of conformal Killing  fields
(CKV) in a flat, three-dimensional space:
\begin{equation} \scrT_k
:=\frac{\partial}{\partial x^k}, \quad \scrS:=x^k
\frac{\partial}{\partial x^k}, \quad \scrR_k:=\epsilon_k
{}^{ij}x_i \frac{\partial}{\partial x^j}, \quad \scrK_k:=x_k
\scrS -\frac{1}{2} r^2 \frac{\partial}{\partial x^k} \, . \end{equation}
The
fields written above correspond\footnote{Number of vectors in
each class is represented by the number in the bracket.} (respectively) to:
 translation (3), scaling (1), rotation (3) and
proper conformal transformation (3). Now we are able to provide
the decomposition of each tensor in basis (\ref{baza}) in a form
given in the Lemma \ref{lemat2}.
\begin{align} 1^\circ \quad &
\scrT_0 \wedge \scrT_k = \scrT_0 \wedge \scrT_k \label{TT}\\ 2^\circ \quad
& \scrT_0 \wedge \scrD = \scrT_0 \wedge \scrS \label{TS}\\ 3^\circ \quad &
\scrT_k \wedge \scrD = -t (\scrT_0 \wedge \scrT_k) - * (\scrT_0
\wedge \scrR_k) \\ 4^\circ \quad & \scrD \wedge \scrL_{0 k}
-\frac{1}{2} \eta(\scrD,\scrD)\scrT_0 \wedge \scrT_k =
-\frac{1}{2} t^2 \scrT_0 \wedge \scrT_k -t * (\scrT_0 \wedge
\scrR_k)+ \scrT_0 \wedge \scrK_k \\ 5^\circ \quad & *(\scrT_0
\wedge \scrT_k) = *(\scrT_0 \wedge \scrT_k) \\ 6^\circ \quad &
*(\scrT_0 \wedge \scrD) = *(\scrT_0 \wedge \scrS) \\ 7^\circ
\quad & *(\scrT_k \wedge \scrD) = -t *(\scrT_0 \wedge \scrT_k) +
(\scrT_0 \wedge \scrR_k) \\ 8^\circ \quad & *(\scrD \wedge
\scrL_{0 k} -\frac{1}{2} \eta(\scrD,\scrD)\scrT_0 \wedge
\scrT_k) = -\frac{1}{2} t^2 *(\scrT_0 \wedge \scrT_k) +t
(\scrT_0 \wedge \scrR_k)+ *(\scrT_0 \wedge \scrK_k)
\end{align}

We have obtained 20 tensors of CYK
basis in a 3+1 decomposition. To calculate the charges, we have
to contract each CYK tensor with a spin-2 field and then
integrate the result over a two-dimensional
surface.
The integration is
done for a fixed moment in time $t$ (formally we would have to
write: for a fixed value of parameter $t$ which enumerates the
leaves of the foliation $\Sigma_t$ in the 3+1 decomposition).
The contraction of a spin-2 field with a CYK tensor (written in
a way proposed in Lemma \ref{lemat2}) reduces to the contraction
of the electric part with a conformal field (for tensors of the
form $\scrT_0\wedge X$) or to the contraction of the magnetic
part with a conformal field (for the tensors of the form
$*(\scrT_0\wedge X)$).

We will write $E(X)$ and $H(X)$ to represent
the charges obtained from a contraction of a conformal Killing
field $X$ with the electric and the magnetic part respectively.
$E_0(X)$ and $H_0(X)$ are the initial values of the
charge, $E_t(X)$ and $H_t(X)$ are the charges in time
\textit{t}. 

Four of the tensors mentioned in  section
\ref{lad} are not time dependent (i.e. tensors with numbers: 1,
2, 5, 6), which also means that the charges associated with
those tensors are constant in time. Four other tensors include
expressions multiplied by  first or  second power of time. Now
we can use the equation (\ref{diwergencja}) and write equations
of the evolution. Let's contract the tensors with numbers 3, 4,
7, 8 from section \ref{lad} with the spin-2 field, then rewrite
the result in terms of the electric and the magnetic part.
Finally we use the equation (\ref{diwergencja}), writing that
this contraction has to be zero, and now we can put all
time-dependent components on  one side of the equation.

\noindent For the charges linear in time we have:
\begin{equation}\label{t1HE} \begin{cases} E_t(\scrR_k)= t\, H(\scrT_k)+
E_0(\scrR_k), \\ H_t(\scrR_k)= -t\, E(\scrT_k)+ H_0(\scrR_k), \\
\end{cases} \end{equation} and for quadratic in time:
\begin{equation}\label{t2HE} \begin{cases} E_t(\scrK_k)=\frac{1}{2}t^2\,
E(\scrT_k) + t\, H(\scrR_k) + E_0(\scrK_k), \\
H_t(\scrK_k)=\frac{1}{2}t^2\, H(\scrT_k) - t\, E(\scrR_k) +
H_0(\scrK_k). \\ \end{cases} \end{equation}

We can classify the
charges according to time-dependence (the number in brackets
enumerates the charges in each class): \begin{itemize} \item (8)
charges constant in time: $E(\scrS)$, $H(\scrS)$, $E(\scrT_k)$,
$H(\scrT_k)$, \item (6) charges linear in time: $E(\scrR_k)$,
$H(\scrR_k)$, \item (6) charges quadratic in time: $E(\scrK_k)$,
$H(\scrK_k)$. \end{itemize} 

\subsection{Partially charged solution -- electric charge counterpart}
\label{rozwiaznieCzesciowoNaladowane}
In \cite{GRG27} we have proposed a ``charged'' spin-2 field configuration which
is a non-oscilating monodipole solution of equation (\ref{rozniczkowy}) with
singularity at $r=0$, and it possesses a global ``potential''
(linearized metric tensor). This is a spin-2 field analog of Coulomb solution
in electrodynamics with electric charge (only).
Let us denote by $\mathbf{p}, \mathbf{k}, \mathbf{s}$ the dipole functions
on a two-dimensional sphere which correspond to constant three-vectors
in Cartesian coordinates $(x^k)$.
More precisely, vector  $p^i$ corresponds to $\mathbf{p}=(p^i x_i)/r$
and, analogously, $\mathbf{s}=(s^i x_i)/r$
and $\mathbf{k}= (k^i x_i)/r$.
The ``charged'' solution  in spherical coordinates
$y^0=t, y^A=(\theta,\varphi), y^3=r$, $A=1,2$
takes the following form:
\begin{equation}
\label{chargedfirst} W_{BC0A} = - \frac{3}{r^2}
\epsilon_{BC}\left(\frac{\mathbf{s}_{,A}}{r} -
\epsilon_A{}^D\mathbf{p}_{,D}\right)\,, \end{equation}
\begin{equation} W_{AB03} = \frac{6}{r^4}\epsilon_{AB}
\mathbf{s}\,, \end{equation} \begin{equation} W_{3A30} =
\frac{3}{r^2}\left(\frac{\epsilon_A{}^D\mathbf{s}_{,D}}{r} +
\mathbf{p}_{_A}\right)\,, \end{equation} \begin{equation} W_{3AB0} =
\frac{3}{r^4} \epsilon_{AB} \mathbf{s}\,, \end{equation}
\begin{equation} W_{3030} = -\frac{2}{r^3}\left(m +
\frac{3\mathbf{k}}{r}\right)\,, \end{equation} \begin{equation}
W_{0A03} = \frac{3}{r^3} \mathbf{k}_{,A}\,, \end{equation}
\begin{equation} W_{ABCD} = \frac{2}{r^3}\left(m +
\frac{3\mathbf{k}}{r}\right)(\eta_{AC} \eta_{BD} - \eta_{AD}
\eta_{BC})\,, \end{equation} \begin{equation} W_{3AB3} =
-W_{0AB0} = \frac{\eta_{AB}}{r^3}\left(m + \frac{3\mathbf{k}}{r}\right)\,,
\end{equation} \begin{equation} \label{chargedlast} W_{BC3A} = -
\frac{3}{r^3} \epsilon_{BC}\epsilon_A{}^D \mathbf{k}_{,D}\,,
\end{equation}
where indices $A,B,C,\dots$ correspond to angular coordinates on $S^2$
and $x^3=r$ is the radial coordinate.\\

According to \cite{GRG27}, spin-2 field solution
(\ref{chargedfirst})--(\ref{chargedlast}) results from linearized metric:
\begin{equation} h_{00} = \frac{2m}{r}+ \frac{2 \mathbf{k}}{r^2}\,,
\end{equation}
\begin{equation} h _{0A} = -6 \mathbf{p} _{,A} - \frac{2}{r}
\epsilon _{A}{} ^{B} \mathbf{s} _{,B}\,, \end{equation}
\begin{equation} h _{03} = - \frac{6 \mathbf{p}}{r}\,,
\end{equation} \begin{equation} h _{33} = \frac{2m}{r}+ \frac{6
\mathbf{k}}{r^2}\,, \end{equation}
which in Cartesian coordinates $(x^k)$ takes the following form:
\begin{equation} h _{00} = \frac{2m}{r}+
\frac{2k_mx^m}{r^2}\,, \end{equation} \begin{equation} h _{0k} =
-\frac{6p_k}{r}- \frac{2}{r^3} \epsilon _{klm}s^lx^m\,,
\end{equation} \begin{equation} h ^{kl} = \frac{x^k x^l}{r^2}
\left( \frac{2m}{r}+ \frac{6k_mx^m}{r^3} \right)\,.
\end{equation}
The above symmetric tensor $h_{\mu\nu}$ is the solution of linearized Einstein
equations with the energy-momentum tensor localized on timelike curve $r=0$
(as a distribution with the support in the center):
\begin{equation} T ^{00} = m \boldsymbol\delta -k^m \boldsymbol\delta _{,m}\,, \quad
T ^{0k} = p^k \boldsymbol\delta +
\frac{1}{2} \epsilon ^{kml}s_l\boldsymbol\delta _{,m}\,,\quad 
T ^{kl}=0\,, \end{equation}
where $\boldsymbol\delta$ denotes three-dimensional Dirac delta ``function'' and
$\epsilon ^{kml}$ is a skew-symmetric Levi-Civita tensor
($ \epsilon ^{123}=1$).

\subsection{Fully charged solution -- magnetic monopole counterpart}
A generalization of (\ref{chargedfirst})--(\ref{chargedlast}), corresponding
in electrodynamics to magnetic monopole, one can obtain by introducing
additional charges $\mathbf{q}= (q^kx_k)/r$, $\mathbf{w} = (w^kx_k)/r$,
$\mathbf{d}= (d^k x_k)/r$.
This is a ``fully charged'' spin-2 field solution which is an analog of
electro-magnetic monopole -- Coulomb solution with electric and magnetic charge.
The quantities $\mathbf{q}$, $\mathbf{w}$, $\mathbf{d}$ are obstructions
for the existence of global linearized metric $h$.

Fully charged spin-2 field solution in spherical coordinates takes the following form
(\cite{GRG27}):
\begin{equation}
\label{firstchargedequation} W_{BC0A} = \epsilon_{BC}
\left(\frac{3}{2r} \mathbf{q}_{,A} +
\frac{3}{r^2}\epsilon_A{}^D\mathbf{p}_{,D} - \frac{3}{r^3}
\mathbf{s}_{,A}\right)\,, \end{equation} \begin{equation} W_{AB03} =
\epsilon_{AB}\left(\frac{3\mathbf{q}}{r^2} + \frac{2b}{r^3} +
\frac{6\mathbf{s}}{r^4}\right)\,, \end{equation} \begin{equation}
W_{3A30} = -\frac{3}{2r}\epsilon_A{}^D\mathbf{q}_{,D} +
\frac{3}{r^2} \mathbf{p}_{,A} + \frac{3}{r^3}\epsilon_A{}^D
\mathbf{s}_{,D}\,, \end{equation} \begin{equation} W_{3AB0} =
\epsilon_{AB}\left(\frac{3\mathbf{q}}{2r^2} +
\frac{b}{r^3}+\frac{3\mathbf{s}}{r^4}\right)\,, \end{equation}
\begin{equation} W_{3003} = \frac{3\mathbf{w}}{r^2} +
\frac{2m}{r^3}+\frac{6\mathbf{k}}{r^4}\,, \end{equation}
\begin{equation} W_{A003} = \frac{3}{2r}\mathbf{w}_{,A} -
\frac{3}{r^3}\mathbf{k}_{,A} -
\frac{3}{r^2}\epsilon_A{}^C\mathbf{d}_{,C}\,, \end{equation}
\begin{equation} W_{ABCD} =\left (\frac{3\mathbf{w}}{r^2} +
\frac{2m}{r^3} +
\frac{6\mathbf{k}}{r^4}\right)\epsilon_{AB}\epsilon_{CD}\,,
\end{equation} \begin{equation}
W_{3AB3}=-W_{0AB0}=\eta_{AB}\left(\frac{3\mathbf{w}}{2r^2}+\frac{m}{r^3}+\frac{3\mathbf{k}}{r^4}\right)\,,
\end{equation} \begin{equation} \label{lastchargedequation}
W_{3ABC}=\epsilon_{BC}\left(\frac{3}{2r}\epsilon_A{}^D\mathbf{w}_{,D}+\frac{3}{r^2}\mathbf{d}_{,A}
-\frac{3}{r^3}\epsilon_A{}^D\mathbf{k}_{,D}\right)\,.
\end{equation}

In Appendix \ref{uzupelnienie}
we give more information about Cartesian form of the formulae
 (\ref{firstchargedequation})--(\ref{lastchargedequation}). 
 Charges $\mathbf{q}$ and $\mathbf{w}$ correspond to metric tensors
 which are not vanishing at spatial infinity ($h _{\mu\nu}=O(1)$).
One can show that $b$ is contained in the metric \begin{equation} h
_{0\phi}=4b \cos\theta\,. \end{equation}
Similarily, charge
 $\mathbf{d}$ with direction 
along $z$-axis ($\mathbf{d}=d\cos\theta$) corresponds to singular metric
\begin{equation} h
_{\theta\phi}=2rd\sin\theta\cos\theta \\ \end{equation} or
\begin{equation} h _{r\theta}= 2d \left( \sin^2\theta \log \tan
\frac{\theta}{2}-\cos\theta \right)\,. \end{equation} 
Definition 
\begin{equation}
	\label{linearE}
E_{kl}:=W_{k00l}\,,
\end{equation}
\begin{equation}
	\label{linearB}
H_{kl}:=\frac{1}{2} W_{0kij}\epsilon^{ij}{}_l
\end{equation}
of electro-magnetic part of spin-2 field applied to the fully
charged solution gives:
\begin{equation}
\begin{aligned}
E_{ij}
=& E(n_i\partial_r+y^A{}_{,i}\partial_A, n_j\partial_r +
y^B{}_{,j}\partial_B) 
= -\frac{\eta_{ij}}{2}\left(\frac{3\mathbf{w}}{r^2} + \frac{2m}{r^3} +
\frac{6\mathbf{k}}{r^4}\right) - \frac{3}{r^2}n^k
\mathbf{d}_{,l}(\epsilon_{kj}{}^ln_i+\epsilon_{ki}{}^ln_j) &\\
& + \frac{3}{2r^2}(n_iw_j + n_j w_i) -
\frac{3}{r^4}(n_ik_j+n_jk_i) 
- n_in_j
\left(-\frac{3\mathbf{w}}{2r^2} - \frac{3m}{r^3} -
\frac{15\mathbf{k}}{r^4}\right),
\end{aligned}
\end{equation}
where $\eta_{ij}=\delta_{ij}$ is the Euclidean metric and $n^k:=x^k/r$,
or in equivalent form:
\begin{equation} \begin{aligned} E_{ij}
=&-\frac{m}{r^3}(\eta_{ij}-3n_in_j) 
-d^l\frac{3}{r^3}n^k(\epsilon_{kjl}n_i+\epsilon_{kil}n_j)+& \\
&-k^l\frac{3}{r^4}(n_l\eta_{ij}+n_i\eta_{lj}+n_j\eta_{li}-5n_in_jn_l) 
-w^l\frac{3}{2r^2}(\eta_{ij}n_l-n_i\eta_{jl}-n_j\eta_{il}-n_in_jn_l)\,,
\end{aligned} \end{equation}
and one more possibility in terms of derivatives of  $1/r$:
\begin{equation} \begin{aligned} E_{ij}=&\quad
m\left(\frac{1}{r}\right)_{,ij}
-\vec{k}\cdot\vec{\nabla}\left(\frac{1}{r}\right)_{,ij}
-[(\vec{d}\times\vec{\nabla})_j\nabla_i+(\vec{d}\times\vec{\nabla})_i\nabla_j)]\frac{1}{r}+
\\
&-w^l\frac{3}{2r^2}(\eta_{ij}n_l-n_i\eta_{jl}-n_j\eta_{il}-n_in_jn_l)\,,
\end{aligned} \end{equation}
and analogously
\begin{equation}
\begin{aligned} H_{ij} =& H(n_i\partial_r+y^A{}_{,i}\partial_A,
n_j\partial_r + y^B{}_{,j}\partial_B) = 
-\frac{\eta_{ij}}{2}\left(\frac{3\mathbf{q}}{r^2} + \frac{2b}{r^3} +
\frac{6\mathbf{s}}{r^4}\right) + \frac{3}{r^2}n^k
\mathbf{p}_{,l}(\epsilon_{kj}{}^ln_i+\epsilon_{ki}{}^ln_j) + &\\
&+\frac{3}{2r^2}(n_iq_j + n_j q_i) -
\frac{3}{r^4}(n_is_j+n_js_i) + 
n_in_j
\left(\frac{3\mathbf{q}}{2r^2} + \frac{3b}{r^3} +
\frac{15\mathbf{s}}{r^4}\right), \end{aligned} \end{equation}
\begin{equation} \begin{aligned}
H_{ij}=&-\frac{b}{r^3}(\eta_{ij}-3n_in_j)  
+p^l\frac{3}{r^3}[n^k(\epsilon_{kjl}n_i+\epsilon_{kil}n_j)]+&
\\
& \hspace{-0.5cm}-s^l\frac{3}{r^4}(n_l\eta_{ij}+n_i\eta_{lj}+n_j\eta_{li}-5n_in_jn_l) 
-q^l\frac{3}{2r^2}(\eta_{ij}n_l-n_i\eta_{jl}-n_j\eta_{il}-n_in_jn_l)\,,
\end{aligned} \end{equation}
and finally in terms of the derivatives of $1/r$:
 \begin{equation} \begin{aligned} H_{ij}=&\quad
\left[\left(b -\vec{s}\cdot\vec{\nabla} \right)\nabla_i\nabla_j
+(\vec{p}\times\vec{\nabla})_j\nabla_i+(\vec{p}\times\vec{\nabla})_i\nabla_j)
\right] \frac{1}{r}+& \\
&-q^l\frac{3}{2r^2}(\eta_{ij}n_l-n_i\eta_{jl}-n_j\eta_{il}-n_in_jn_l)\,.
\end{aligned} \end{equation}
Let us observe that exchanging  $w\rightarrow q$, $k\rightarrow
s$, $d \rightarrow -p$ and $m \rightarrow b$ in the electric part $E$
we get magnetic part $H$.
$E(w\rightarrow q, k\rightarrow s,
d \rightarrow -p, m \rightarrow b) = H$
represents spin-2 field version of electromagnetic symmetry between electric and magnetic monopole.

Finally, we can check the values of quasi-local charges for the ``fully charged'' solution:

\begin{equation}
\label{firstintegral}
E(\mathcal{S})=Q(E,\mathcal{S})=\int\limits_{S(r)}E_{ij}\mathcal{S}^jn^i dS =
8\pi m\,, \end{equation} \begin{equation}
E(\mathcal{T}_k)=Q(E,\mathcal{T}_k)=\int\limits_{S(r)}E_{ij}\mathcal{T}^j_kn^i
dS=8\pi w_k\,, \end{equation} \begin{equation}
E(\mathcal{R}_k)=Q(E,\mathcal{R}_k)=\int\limits_{S(r)}E_{ij}\mathcal{R}^j_kn^i
dS= -8\pi d_k\,, \end{equation} \begin{equation}
E(\mathcal{K}_k)=Q(E,\mathcal{K}_k)=\int\limits_{S(r)}E_{ij}\mathcal{K}^j_kn^i
dS= 8\pi k_k\,, \end{equation} \begin{equation}
H(\mathcal{S})=Q(H,\mathcal{S})=\int\limits_{S(r)}H_{ij}\mathcal{S}^jn^i dS =
8\pi b\,, \end{equation} \begin{equation}
H(\mathcal{T}_k)=Q(H,\mathcal{T}_k)=\int\limits_{S(r)}H_{ij}\mathcal{T}^j_kn^i
dS= 8\pi q_k\,, \end{equation} \begin{equation}
H(\mathcal{R}_k)=Q(H,\mathcal{R}_k)=\int\limits_{S(r)}H_{ij}\mathcal{R}^j_kn^i
dS= 8\pi p_k\,, \end{equation} \begin{equation}
\label{lastintegral}
H(\mathcal{K}_k)=Q(H,\mathcal{K}_k)=\int\limits_{S(r)}H_{ij}\mathcal{K}^j_kn^i
dS= 8\pi s_k\,. \end{equation}
CYK tensors (\ref{TT}) and (\ref{TS}) correspond to four time-independent charges:
\begin{equation}
	\label{firstDynamic}
	\dot{Q}(E, \mathcal{T}_k) = \dot{Q}(E, \mathcal{S}) = \dot{Q}(H,
	\mathcal{T}_k) = \dot{Q}(H, \mathcal{S}) =0 \,.
\end{equation}
Moreover, from (\ref{t1HE}) and (\ref{t2HE}) we obtain time-dependence of other quantities:
\begin{equation}
	\dot{Q}(E, \mathcal{R}_k) = Q(H, \mathcal{T}_k) \,,
\end{equation}
\begin{equation}
	\dot{Q}(E, \mathcal{K}_k) = Q(H, \mathcal{R}_k) \,,
\end{equation}
\begin{equation}
	\dot{Q}(H, \mathcal{R}_k) = -Q(E, \mathcal{T}_k) \,,
\end{equation}
\begin{equation}
	\label{lastDynamic}
	\dot{Q}(H, \mathcal{K}_k) = -Q(E, \mathcal{R}_k) \,.
\end{equation}
Finally, we have the following time evolution for the charges:
eight quantities are constant, six of them are linear and other six are quadratic in time.
More precisely, we have constant charges:
\begin{equation} m(t)=m(0), \quad
w_l(t)=w_{l}(0) ,\quad b(t)=b(0) ,\quad q_l(t)=q_{l}(0)\,,
\end{equation}
linear in time:
\begin{equation} p_l(t) =
-tw_{l}(0) + p_{l}(0), \qquad d_l(t) = -tq_{l}(0) + d_{l}(0)\,,
\end{equation}
and quadratic in time:
\begin{equation} k_l(t) =
-\frac{1}{2}t^2w_{l}(0) + tp_{l}(0) +k_{l}(0)\,, 
\qquad s_l(t) = -\frac{1}{2}t^2q_{l}(0) +td_{l}(0)
+s_{l}(0)\,, \end{equation}
where $m(0)$ denotes initial value of $m$ at $t=0$ and similarly for the rest of the quantities.

\paragraph{Traditional Poincar\'e charges}
Let us observe that traditional relations between angular momentum
(or center of mass) and Killing vectors (e.g. ADM or Komar formula)
are substituted by conformal acceleration. More precisely, we have the following table:\\[3ex]
\centerline{
\begin{tabular}{|c|c|c|cl|}
\hline
KV & Charges & CKV & & \\
\hline
$\mathcal{T}_0$ & $p_0 \leftrightarrow m$  & $\mathcal{S}$ & (1) & energy \\
$\mathcal{T}_k$ & $p_k \leftrightarrow {\bf p}$ & $\mathcal{R}_k$ & (3) & linear momentum \\
$\mathcal{L}_{kl}$ & $j_{kl} \leftrightarrow {\bf s}$ & $\mathcal{K}_k$ & (3) & angular momentum \\
$\mathcal{L}_{0k}$ & $j_{0k} \leftrightarrow {\bf k}$ & $\mathcal{K}_k$ & (3) & center of mass \\
\hline
\end{tabular}
}
\\[3ex]
Other quantities: ${\bf b}$ -- dual mass, ${\bf d}$ -- dual momentum ,
${\bf w}$ -- linear acceleration, ${\bf q}$ -- angular acceleration
are usually vanishing, if we want to have global ``potentials''
(linearized metric $h$ like vector potential $A$ for magnetic monopole).
However, some parameters in Einstein metrics can be interpreted as topological charges, e.g. dual mass appears in Taub-NUT
 solution (\cite{JJMLnut},\cite{JJcykads}) and dual momentum in Demia\'nski metrics (\cite{GRG27},\cite{GP}).
In \cite{GP} a large class of metric tensors is given (see also eq. (4.50) in \cite{JJcykads}).
It would be nice to check, if some parameters in those spacetimes correspond to charges ${\bf q}$ and ${\bf w}$ in some asymptotic regime.

\section{Schwarzschild spacetime}
\paragraph{Introduction} \label{subSch} For the
Schwarzschild metric the construction described above for the
Minkowski spacetime cannot be repeated, because the equation
defining CYK tensors ($\scrQ_{\lambda \kappa \sigma}=0$) has
only two solutions (for Minkowski we had 20). One of them
corresponds to the mass (i.e. after calculating the
integral we get the charge corresponding to the mass which is
equal to the parameter $M$ appearing in a standard form of the
Schwarzschild metric). The second solution can be classified as
a dual mass and in our case it vanishes. Therefore we are forced
to apply a different construction. Our goal is to define charges
which are ``local'' in time.

Let's begin with writing down the Schwarzschild metric
in the parametrization ($t$, $r$, $\theta$, $\phi$):
\begin{equation} g_{\mu \nu} dx^\mu
dx^\nu=-\left(1-\frac{2M}{r}\right)dt^2+\frac{dr^2}{1-\frac{2M}{r}}+r^2\left(d\theta^2+\sin^2\theta
d\phi^2\right). \end{equation} Now let us introduce a new
coordinate $\bar{r}$ defined by the equality
$r=\bar{r}(1+\frac{M}{2\bar{r}})^2$. 
Schwarzschild metric in coordinates ($t$, $\bar{r}$, $\theta$,
$\phi$) takes the form: \begin{equation} g_{\mu \nu} dx^\mu
dx^\nu=-\left(\frac{1-M/2\bar{r}}{{1+M/2\bar{r}}} \right)^2 dt^2
+\left(1+\frac{M}{2\bar{r}} \right)^4 \left[d
\bar{r}\:{}^2+\bar{r}\:{}^2 \left(d\theta^2+\sin^2\theta
d\phi^2\right)\right]\,. \label{newschwarz} \end{equation} 
We observe that for a fixed value of $t$,
the metric (\ref{newschwarz}) is conformally flat (i.e. it takes
the form of a flat-space metric multiplied by a conformal
factor $\left(1+\frac{M}{2\bar{r}} \right)^4$). 
\noindent Using this fact we conclude that we have a set of ten
conformal Killing fields, which are identical to the ones that
we had for a flat 3-D space, since both metrics are conformally
equivalent.  Moreover, using the formula
$K_{ik}=\frac{1}{2N}(N_{i|k}+N_{k|i}-\frac{\partial
g_{ik}}{\partial t})$ we can easily check that the extrinsic
curvature tensor $K_{ik}$ vanishes. 
\paragraph{Definition of momentary charges} In the
previous sections we have used the contraction of a spin-2
field with CYK tensors to define global charges, then we have
shown (Lemma 2) that CYK tensors can be expressed as
contractions of electric (or magnetic) part with conformal
Killing fields. Following this lead we will try to provide a
definition of the momentary charges for the Schwarzschild metric
as contractions of $E$ and $H$ with conformal fields (ignoring
the fact that we  have not enough CYK tensors for this metric).

Let $E_{kl}$ be an electric part, $H_{kl}$ a magnetic part and
$X^l$ a conformal Killing field, then \begin{equation} \begin{split}
(E^k{}_l
X^l){}_{|k}&=(E^{kl}X_l){}_{|k}=E^{kl}{}_{|k}X_l+E^{kl}X_{l|k}=E^{kl}{}_{|k}X_l+E^{kl}X_{(l|k)}=\\&=E^{kl}{}_{|k}X_l+\lambda
E^{kl}g_{kl}=E^{kl}{}_{|k}X_l\,, \end{split} \end{equation} where
we have used the fact that $E^{kl}$ is a symmetric traceless
tensor. Identical calculation can be repeated for the magnetic
part $H^{kl}$.

Now we have to provide an expression for a
three-dimensional divergence of an electric and a magnetic
part. Let's use the following formula: \begin{equation}
(g^{\sigma \mu} W_{\sigma \lambda \kappa \nu} n^\lambda
n^\nu)_{;\mu}=(W^\mu {}_{\lambda \kappa \nu})_{; \mu} n^\lambda
n^\nu + g^{\sigma \mu} W_{\sigma \lambda \kappa \nu} (n^\lambda
n^\nu)_{; \mu}\,. \label{1} \end{equation} The first component on
the right-hand-side is zero because of one of the properties
of the tensor $W$  (i.e. the four-dimensional divergence of
$W$  is zero iff (\ref{rozniczkowy}), from the definition
of spin-2 field). Now let's work on the left-hand-side: \begin{equation}
(g^{\sigma \mu} W_{\sigma \lambda \kappa \nu} n^\lambda
n^\nu)_{;\mu}=((g^{\sigma \mu} + n^\sigma n^\mu) W_{\sigma
\lambda \kappa \nu} n^\lambda n^\nu)_{;\mu}=-(\overset{3}{g}{}^{i
j} E_{j \kappa})_{;i}\,. \end{equation} In the first equality we
have added $n^\sigma n^\mu$, using the fact that $W$ is
antisymmetric, and then we applied the definition of the electric
part. \begin{equation} (\overset{3}{g}{}^{i j} E_{j
\kappa})_{;i}=(\sqrt{-g}E^i{}_\kappa)_{;i}\frac{1}{\sqrt{-g}}=
(N\sqrt{\overset{3}{g}}E^i{}_\kappa)_{|i}\frac{1}{N\sqrt{\overset{3}{g}}}=E^i{}_{\kappa|i}+E^i{}_\kappa
(\log N)_{,i}\,. \label{2} \end{equation} We managed to transform
left-hand-side of (\ref{1}) to the form containing three-dimensional
covariant divergence of the electric part. In equality (\ref{2})
we used the formula
$(\sqrt{-g}E^i{}_\kappa)_{;i}=(N\sqrt{\overset{3}{g}}E^i{}_\kappa)_{|i}$,
which is proven in the appendix to this paper. Right-hand-side of (\ref{1})
was calculated in a straightforward way for the metric
(\ref{newschwarz}). The details of this calculation are
presented in the appendix. Finally we derived the following
formula: \begin{equation} E^i{}_{j|i}=0, \label{divE}
\end{equation} and from an identical reasoning for the magnetic
part we have \begin{equation} H^i{}_{j|i}=0. \end{equation} 

Both three-dimensional covariant divergences are zero. This
means that the momentary gravitational charges (defined as a
contraction of an electric or a magnetic part with a conformal
Killing field) fulfill the Gauss law without sources. We can
conclude that the value of the charge would not depend on the
choice of the integration surface if there exists an appropriate
volume just like it was for the Minkowski spacetime. This means
that if we want to calculate the total charge inside any closed
surface, we don't need to know the configuration of fields
$E_{ab}$ and $H_{ab}$ inside this surface, we only need to know
the values at the boundary. 
Results
presented in this section (i.e. that the three-dimensional
covariant divergence of an electric and a magnetic part is zero)
could also be reproduced using formulas described in paper
\cite{anderson}: \begin{equation} E^i{}_{j|i}=+(K \wedge H){}_j\,,
\label{monE} \end{equation} \begin{equation} H^i{}_{j|i}=-(K
\wedge E){}_j\,, \label{monH} \end{equation} where $\wedge$ is the
 operation defined below for two symmetric tensors:
\begin{equation} (A \wedge B){}_a=\epsilon_a{}^{bc}A_b{}^d
B_{dc}\,. \end{equation} Equations (\ref{monE}) and (\ref{monH})
imply that this construction of momentary charges can be applied
to any metric fulfilling following conditions: \begin{itemize}
\item spatial part of the metric is conformally flat for  fixed
time\footnote{Because we need all ten CKVs.}, \item extrinsic
curvature tensor vanishes. \end{itemize}

\section{Conclusions} In this paper we reconsider the definition of quasi-local
gravitational charges for the
Minkowski spacetime in terms of conformal Yano-Killing tensors and a
spin-2 field. The set of twenty charges, defined in that way,
has properties very similar to the ones that are also valid for
the electric charge, for example our gravitational charges
fulfill Gauss law.
We extend the definition of quasi-local
gravitational charges.
The Lemma describing the decomposition of CYK
tensors into exterior product of a time translation vector field and a conformal Killing
vector has lead us to the idea of defining momentary charges (as
a contraction of an electric and a magnetic part with the
conformal field) for a wide range of metrics (e.g. when the spatial
part is conformally flat for fixed time and the
extrinsic curvature vanishes), including the
Schwarzschild spacetime. Moreover, we have proven by
straightforward calculation that the charges, defined using the
described procedure, fulfill Gauss law and we have given the
conditions that are satisfactory to repeat this construction for
a well-defined class of metrics.

In the future we would like to apply this construction for the case of asymptotically flat initial data.
It is well known that some spacetimes admit (exact) CYK tensors
(\cite{cykem},\cite{JJMLKerr},\cite{JJMLnut},\cite{JJcykads}) but in general
one should consider asymptotic CYK tensors (\cite{kerrnut}) which correspond
to the notion of strong asymptotic flatness. The existence of asymptotic
conformal Killing vectors is less restrictive and it should lead to the
definition of global momentary charge for different asymptotics at spatial infinity.
In particular, angular momentum and center of mass correspond to conformal acceleration $\mathcal{K}$.

The content of this paper is the answer to the following question:\\
{\em What is the analog of Coulomb solution (electric and magnetic monopole) for the spin-2 field?}\\
For spin-1 field the solution is ``monopole'', for spin-2 field we have also dipole part.
In Maxwell theory we have only time independent charges, for gravity we get also time-dependent quantities.
The ``wave part'' of the theory starts from dipoles ($l=1$) for electrodynamics
and respectively quadrupoles ($l=2$) for gravity.
Hence the ``charged part'' for spin-1 field is represented by $l=0$ but for spin-2 field we have $l=0$ and $l=1$.
Finally, the analog of the electric-magnetic monopole in electrodynamics is given by the mono-dipole solution
(\ref{chargedfirst})--(\ref{chargedlast}) for spin-2 field.\\[2ex]
{\noindent \sc Acknowledgements}  Supported in part by Narodowe Centrum Nauki under the grant DEC-2011/03/B/ST1/02625.

\appendix

\section{Fully charged solution}
\label{uzupelnienie}
Cartesian components:
\begin{equation}
	W_{0ij0}=\frac{3m}{r^3}n_i n_j+\frac{15 \mathbf{k}}{r^4}n_i n_j -
	\frac{3}{r^4}(k_jn_i + k_i n_j)
	-\frac{\eta_{ij}}{r^3}\left(m+\frac{3\mathbf{k}}{r}\right)\,,
\end{equation}
\begin{equation}
	\begin{aligned}
		W_{ijkl}=&\frac{3}{r^3}\left(m+\frac{3\mathbf{k}}{r}\right)(n_i n_l \eta_{jk} - n_i n_k
		\eta_{jl} + n_j n_k \eta_{il} - n_j n_l \eta_{ik}) + 
\frac{2}{r^3}\left(m+\frac{3\mathbf{k}}{r}\right)(\eta_{ik} \eta_{jl} -
		\eta_{il}\eta_{jk}) + \\
		+& \frac{3}{r^3}\epsilon_{mkl}n^m n^p k_{,h}(\epsilon_{pi}{}^h n_j -
		\epsilon_{pj}{}^h n_i) + 
\frac{3}{r^3}\epsilon_{mij}n^m n^p k_{,h}(\epsilon_{pk}{}^h n_l -
		\epsilon_{pl}{}^h n_k)\,,
	\end{aligned}
\end{equation}
\begin{equation}
	\begin{aligned}
		W_{0ijk}=& \frac{3}{r^3}[(p_j-\mathbf{p} n_j) n_i n_k - (p_k-\mathbf{p}
		n_k)n_i n_j] + 
\frac{3}{r^4}( n_i n_k n^m \epsilon_{mj}{}^l s_l - n_i n_j n^m
		\epsilon_{mk}{}^l s_l) + \\
		+&\frac{3}{r^4} n^m \mathbf{s}(3\epsilon_{mjk}n_i +\epsilon_{mik} n_j +
		\epsilon_{mji} n_k) - 
\frac{3}{r^4}\epsilon_{mjk}n^m s_i + 
\frac{3}{r^3}\epsilon_{mjk} n^m n^h \epsilon_{hi}{}^l p_l\,.
	\end{aligned}
\end{equation}
Contraction with CKV:
	$\mathcal{T}_k=\frac{\partial}{\partial x^k}$,
	$\mathcal{S}=x^k \frac{\partial}{\partial x^k}$,
	$\mathcal{R}_k=\epsilon_k{}^{ij}x_i\frac{\partial}{\partial x^j}$,
	$\mathcal{K}_k=x_k\mathcal{S}-\frac{1}{2}r^2\frac{\partial}{\partial x^k}$,
gives
	\begin{equation}
		\begin{aligned}
			E _{ik}\mathcal{T}^k_j &= -\frac{\eta_{ij}}{2}\left(\frac{3\mathbf{w}}{r^2} +
		\frac{2m}{r^3} + \frac{6\mathbf{k}}{r^4}\right) - \frac{3}{r^2}n^k
		\mathbf{d}_{,l}(\epsilon_{kj}{}^ln_i+\epsilon_{ki}{}^ln_j) +
		\\&+
\frac{3}{2r^2}(n_iw_j + n_j w_i) - \frac{3}{r^4}(n_ik_j+n_jk_i) 
-n_in_j
		\left(-\frac{3\mathbf{w}}{2r^2} - \frac{3m}{r^3} - \frac{15\mathbf{k}}{r^4}\right)\,,
		\end{aligned}
	\end{equation}
	\begin{equation}
		\begin{aligned}
		E^i{}_j\mathcal{R}_k{}^j=-\epsilon_k{}^{li}n_l\left(\frac{3\mathbf{w}}{2r} +
		\frac{m}{r^2}+\frac{3\mathbf{k}}{r^3}\right)-\frac{3}{r}n^i\mathbf{d}_{,k}-
		\epsilon_k{}^{mj}x_m\left(\frac{3}{r^4}n^ik_j-\frac{3}{2r^2}n^iw_j\right)\,,
		\end{aligned}
	\end{equation}
	\begin{equation}
		\begin{aligned}
			E_{ij}\mathcal{K}^j_k &=\frac{m}{2r}(n_in_k+\eta_{ik})-d^l\frac{3}{2r}n^m
		(\epsilon_{mil}n_k-\epsilon_{mkl}n_i)+\\
		&-k^l\frac{3}{2r^2}(-n_ln_kn_i+n_k\eta_{li}-n_l\eta_{ik}-n_i\eta_{lk})+\\
		&-w^l\frac{3}{4}(-n_in_kn_l-n_k\eta_{li}+n_i\eta_{kl}-n_l\eta_{ik})\,,
		\end{aligned}
	\end{equation}
	\begin{equation}
		\begin{aligned}
		E^i{}_j S^j =
		n^i\left(\frac{2m}{r^2}+\frac{9\mathbf{k}}{r^3}\right)-\frac{3}{r}n^k\mathbf{d}_{,l}
		\epsilon_k{}^{il}+\frac{3}{2r}(n^i\mathbf{w}+w^i)-\frac{3}{r^3}k^i\,,
		\end{aligned}
	\end{equation}
	\begin{equation}
		\begin{aligned}
			H _{ik}\mathcal{T}^k_j &= -\frac{\eta_{ij}}{2}\left(\frac{3\mathbf{q}}{r^2} +
		\frac{2b}{r^3} + \frac{6\mathbf{s}}{r^4}\right) + \frac{3}{r^2}n^k
		\mathbf{p}_{,l}(\epsilon_{kj}{}^ln_i+\epsilon_{ki}{}^ln_j) +
		\\&+\frac{3}{2r^2}(n_iq_j + n_j q_i) - \frac{3}{r^4}(n_is_j+n_js_i) 
+n_in_j
		\left(\frac{3\mathbf{q}}{2r^2} + \frac{3b}{r^3} + \frac{15\mathbf{s}}{r^4}\right)\,,
		\end{aligned}
	\end{equation}
	\begin{equation}
		\begin{aligned}
		H^i{}_j\mathcal{R}_k{}^j=\epsilon_k{}^{li}n_l\left(-\frac{3\mathbf{q}}{2r} -
		\frac{b}{r^2}-\frac{3\mathbf{s}}{r^3}\right)+\frac{3}{r}n^i\mathbf{p}_{,k}+
		\epsilon_k{}^{mj}x_m\left(-\frac{3}{r^4}n^is_j+\frac{3}{2r^2}n^iq_j\right)\,,
		\end{aligned}
	\end{equation}
	\begin{equation}
		\begin{aligned}
			H_{ij}\mathcal{K}^j_k &=\frac{b}{2r}(n_in_k+\eta_{ik})+p^l\frac{3}{2r}n^m
		(\epsilon_{mil}n_k-\epsilon_{mkl}n_i)+\\
		&-s^l\frac{3}{2r^2}(-n_ln_kn_i+n_k\eta_{li}-n_l\eta_{ik}-n_i\eta_{lk})+\\
		&-q^l\frac{3}{4}(-n_in_kn_l-n_k\eta_{li}+n_i\eta_{kl}-n_l\eta_{ik})\,,
		\end{aligned}
	\end{equation}
	\begin{equation}
		\begin{aligned}
		H^i{}_j S^j =
		n^i\left(\frac{2b}{r^2}+\frac{9\mathbf{s}}{r^3}\right)+\frac{3}{r}n^k\mathbf{p}_{,l}
		\epsilon_k{}^{il}+\frac{3}{2r}(n^i\mathbf{q}+q^i)-\frac{3}{r^3}s^i\,.
		\end{aligned}
	\end{equation}
After contraction with normal we get:
\begin{equation}
	\label{dodatkowezN1}
	E_{ij}\mathcal{T}^j_kn^i=\frac{2m}{r^3}n_k-\frac{3}{r^3}d^ln^m\epsilon_{mkl}
	-\frac{3}{r^4}k_k+\frac{9}{r^4}k^ln_ln_k+\frac{3}{2r^2}(w_k+w^ln_ln_k)\,,
\end{equation}
\begin{equation}
	E_{ij}\mathcal{R}^j_kn^i=-\frac{3}{r^2}(d^k-d^ln_ln^k)-\frac{3}{r^3}k^ln_p
	\epsilon^p{}_{lk}+\frac{3}{2r}w^ln_p\epsilon^p{}_{lk}\,,
\end{equation}
\begin{equation}
	E_{ij}\mathcal{K}^j_kn^i=\frac{m}{r}n_k+\frac{3}{2r}d^ln^m\epsilon_{mkl}+
	\frac{3}{2r^2}k^l(n_ln_k+\eta_{kl})-\frac{3}{4}w^l(\eta_{kl}-3n_kn_l)\,,
\end{equation}
\begin{equation}
	E_{ij}\mathcal{S}^jn^i=\frac{2m}{r^2}+\frac{6}{r^3}k_ln^l+\frac{3}{r}w_ln^l\,,
\end{equation}
\begin{equation}
	H_{ij}\mathcal{T}^j_kn^i=\frac{2b}{r^3}n_k+\frac{3}{r^3}p^ln^m\epsilon_{mkl}
	-\frac{3}{r^4}s_k+\frac{9}{r^4}s^ln_ln_k+\frac{3}{2r^2}(q_k+q^ln_ln_k)\,,
\end{equation}
\begin{equation}
	H_{ij}\mathcal{R}^j_kn^i=\frac{3}{r^2}(p^k-p^ln_ln^k)-\frac{3}{r^3}s^ln_p
	\epsilon^p{}_{lk}+\frac{3}{2r}q^ln_p\epsilon^p{}_{lk}\,,
\end{equation}
\begin{equation}
	H_{ij}\mathcal{K}^j_kn^i=\frac{b}{r}n_k-\frac{3}{2r}p^ln^m\epsilon_{mkl}+
	\frac{3}{2r^2}s^l(n_ln_k+\eta_{kl})-\frac{3}{4}q^l(\eta_{kl}-3n_kn_l)\,,
\end{equation}
\begin{equation}
	\label{dodatkowezN2}
	H_{ij}\mathcal{S}^jn^i=\frac{2b}{r^2}+\frac{6}{r^3}s_ln^l+\frac{3}{r}q_ln^l\,.
\end{equation}
Integrating expressions
(\ref{dodatkowezN1})--(\ref{dodatkowezN2}) on $S^2$
we obtain (\ref{firstintegral})--(\ref{lastintegral}).

 \begin{subsection}*{Proof of the lemma
\ref{lemat1}} \[ \nabla _\nu F^{\mu \nu}(W,Q)=\nabla_\nu (W^{\mu
\nu \alpha \beta} Q_{\alpha \beta})=(\nabla_\nu W ^{\mu \nu
\alpha \beta})Q_{\alpha \beta} + W^{\mu \nu \alpha
\beta}(\nabla_\nu Q_{\alpha \beta}) \] The first component is
zero because of the equation (\ref{11}), for the second
component: \[ \begin{split} W^{\mu \nu \alpha
\beta}\scrQ_{\alpha\beta\nu}&=W^{\mu \nu \alpha
\beta}\left[Q_{\lambda \kappa ; \sigma}+Q_{\sigma \kappa ; \lambda}-
\frac{2}{3}(g_{\sigma \lambda} {Q^\nu}_{\kappa ;\nu}+ g_{\kappa
( \lambda} Q_{\sigma )}{}^\mu{}_{;\mu})\right]=\\ &=W^{\mu \nu \alpha
\beta}(Q_{\alpha \beta;\nu}+Q_{\nu \beta ; \alpha})=(W^{\mu \nu
\alpha \beta}+W^{\mu \alpha \nu \beta})Q_{\alpha \beta ;\nu}=\\
&= \left(W^{\mu \nu \alpha \beta}+\frac{1}{2}W^{\mu \alpha \nu
\beta}-\frac{1}{2}W^{\mu \beta \nu \alpha}\right)Q_{\alpha \beta
;\nu}=\\&=\left(W^{\mu \nu \alpha \beta}+\frac{1}{2}W^{\mu \alpha \nu
\beta}+\frac{1}{2}W^{\mu \beta \alpha \nu}\right)Q_{\alpha \beta
;\nu}=\\&=\frac{3}{2}W^{\mu \nu \alpha \beta}Q_{\alpha
\beta;\nu}\,. \end{split} \] In the first line we used the fact
that $W$ is traceless, in the second we renamed the indices, in
the third we observed that $Q$ is antisymmetric, and in the
fourth line we applied equality $W^{\mu [\nu \alpha
\beta]}=0$.\\ Finally we get: \[ \nabla _\nu F^{\mu
\nu}(W,Q)=\frac{2}{3}W^{\mu \nu \alpha \beta} \scrQ_{\alpha
\beta \nu}. \] \end{subsection}
\begin{subsection}*{Divergence of a tensor density with two
indices} Let $\scrL^i {}_j$ be a tensor density, so \[ \scrL^i
{}_{j;i}=\sqrt{g} (\frac{1}{\sqrt{g}} \scrL ^i
{}_j)_{;i}=\sqrt{g} (L^i{}_j)_{;i}=\sqrt{g}[L^i{}_{j,i}+L^k{}_j
\Gamma ^i{}_{k i}-L^i{}_k \Gamma^k{}_{j i}]= \] \[ =\sqrt{g}
L^i{}_{j,i}+L^i{}_j \sqrt{g}_{,i}-\sqrt{g}L^i{}_k
\Gamma^k{}_{ji}=(\scrL^i{}_j)_{,i}-\sqrt{g}L^i{}_k
\Gamma^k{}_{ji}\,. \] Assume that $L^{ik}$ is a symmetric tensor
and perform the contraction with a Christoffel symbol: \[
L^i{}_k \Gamma^k{}_{ji}=\frac{1}{2} L^{i
k}(g_{kj,i}-g_{ji,k}+g_{ik,j})=\frac{1}{2} L^{i k}g_{ik,j}\,. \]
Finally for a symmetric tensor $L^{ik}$ we get: \begin{equation}
\scrL^i {}_{j;i}=(\scrL^i{}_j)_{,i}-\frac{1}{2} \sqrt{g}
L^{ik}g_{ik,j}=(\scrL^i{}_j)_{,i}-\frac{1}{2} \scrL^{ik}g_{ik,j}\,.
\label{3} \end{equation}
Now we can prove the formula
$(\sqrt{-g}E^i{}_\kappa)_{;i}=(N\sqrt{\overset{3}{g}}E^i{}_\kappa)_{|i}$
used in equality (\ref{2}). Let $\scrE_{\mu \nu}=\sqrt{-g}W_{\mu
\lambda \kappa \nu}n^\lambda n^\kappa$. Therefore we can write:
\[ \scrE^\mu{}_k=g^{\mu \lambda}\scrE_{\lambda k}=(g^{\mu
\lambda}+n^\mu n^\lambda)\scrE_{\lambda
k}=\overset{3}{g}\;{}^{\mu \lambda}\scrE_{\lambda k}\,. \] Next
using the equation (\ref{3}) we obtain: \begin{equation} \begin{split}
\scrE^\mu{}_{k ; \mu}&=\scrE^\mu{}_{k,\mu}-\frac{1}{2}\scrE^{\mu
\nu}g_{\mu \nu, k}=\scrE^i{}_{k,i}-\frac{1}{2}\scrE^{i j }g_{i
j,k}=\\ & (N
\sqrt{\overset{3}{g}}E^i{}_k)_{,i}-\frac{1}{2}N\sqrt{\overset{3}{g}}E^{ij}g_{ij,k}
=(N\sqrt{\overset{3}{g}}E^i{}_k)_{|i}\,.
\end{split} \end{equation} \end{subsection}
\begin{subsection}*{Right-hand-side of (\ref{1}) for the metric
(\ref{newschwarz})} \begin{equation} \begin{split}
\hspace*{-0.5cm} g^{\sigma
\mu}W_{\sigma \lambda \kappa \nu} (n^\lambda
n^\nu){}_{;\mu}&
=g^{\sigma \mu}W_{\sigma \lambda \kappa \nu}
(n^\lambda n^\nu){}_{,\mu}+g^{\sigma \mu}W_{\sigma \lambda
\kappa \nu}n^\lambda n^\alpha \Gamma^\nu_{\alpha \mu} +g^{\sigma
\mu}W_{\sigma \lambda \kappa \nu} n^\nu n^\alpha
\Gamma^\lambda_{\alpha \mu} \\ &=g^{11}W_{10\kappa
0}\left(\frac{1}{N^2}\right){}\!\!_{,1}+g^{\sigma \mu}W_{\sigma 0 \kappa \nu}
\frac{1}{N^2} \Gamma^\nu_{0\mu}+g^{\sigma \mu} W_{\sigma \lambda
\kappa 0}
\frac{1}{N^2}\Gamma^\lambda_{0\nu} \\&=g^{11}W_{10\kappa
0}\left(\frac{1}{N^2}\right){}\!\!_{,1}
+\frac{1}{N^2}\Gamma^0_{01}g^{11}W_{10\kappa
0}+\frac{1}{N^2}(\Gamma^0_{01}g^{11}-g^{00}\Gamma^1_{00})W_{10
\kappa 0} \\&=\frac{1}{N^2}W_{10\kappa
0}[2\Gamma^0_{01}g^{11}-g^{00}\Gamma^1_{00}-2g^{11}(\log(N)){}_{,1}]
\end{split} \label{rach} \end{equation} In the calculation above
we used the fact that each Christoffel symbol of the form
$\Gamma^\mu_{0 \nu}$ excluding $\Gamma^0_{01}$ and
$\Gamma^1_{00}$ is zero. \end{subsection}
\begin{subsection}*{Proof of the formula (\ref{divE})} To derive
the final formula for the divergence of the electric part, we
have to use the results from equations (\ref{2}) and
(\ref{rach}): \begin{equation} E^i{}_{\kappa |
i}=-\frac{1}{N^2}W_{10\kappa
0}[2\Gamma^0_{01}g^{11}-g^{00}\Gamma^1_{00}-3g^{11}\log(N){}_{,1}].
\end{equation} After substituting the values for the metric
(\ref{newschwarz}) we obtain \begin{equation} E^i{}_{\kappa | i}=0.
\end{equation} \end{subsection}
\begin{subsection}*{Summary of
basic facts used in a 3+1 decomposition}
Four-dimensional
metric can be expressed using the lapse ($N$) and the shift
($N^m$) in a following manner:
\[ \begin{bmatrix}
g_{00}&g_{0k}\\g_{i0}&g_{ik} \end{bmatrix}=\begin{bmatrix} (N_s
N^s-N^2)& N_k \\ N_i & g_{ik} \end{bmatrix}. \] Inverse 
metric: \[ \begin{bmatrix} g^{00}&g^{0m}\\g^{k0}&g^{km}
\end{bmatrix}=\begin{bmatrix} -(1/N^2)& (N^m/N^2)\\ (N^k/N^2) &
(g^{km}-N^kN^m/N^2) \end{bmatrix}. \]
Unit normal time-like
vector:
\[ n_\mu=(-N,0,0,0), \qquad n^\mu=[(1/N),-(N^m/N)]. \]
Connection of three-dimensional metric with a four-dimensional
one:
\[ \overset{4}{g}{}^{\mu \nu}+n^\nu n^\nu =
\overset{3}{g}{}^{\mu \nu}, \qquad \overset{4}{g}{}_{\mu
\nu}+n_\nu n_\nu = \overset{3}{g}{}_{\mu \nu}. \] Volume element:
\[
(-\overset{4}{g}){}^{1/2}\textrm{dx}^0\textrm{dx}^1\textrm{dx}^2\textrm{dx}^3=N\;
\overset{3}{g}\;{}^{1/2}\textrm{dt}\;\textrm{dx}^1\textrm{dx}^2\textrm{dx}^3.
\] \end{subsection}
\begin{subsection}*{Non-zero Christoffel symbols of metric
(\ref{newschwarz})} \[ \begin{split}
\Gamma^0_{01}&=-\frac{4M}{M^2-4 \bar{r}{}^2},   \qquad
\Gamma^1_{00}=-\frac{64M\bar{r}{}^4(M-2\bar{r})}{(M+2\bar{r})^7}, \qquad
\Gamma^1_{11}=-\frac{2M}{\bar{r}(M+2\bar{r})} & \\
\Gamma^1_{22}&=\frac{\bar{r}(M-2\bar{r})}{M+2\bar{r}},   \qquad
\Gamma^1_{33}=\frac{\bar{r}\sin^2\theta(M-2\bar{r})}{M+2\bar{r}},  \qquad
\Gamma^2_{12}=-\frac{M-2\bar{r}}{\bar{r}(M+2\bar{r})}&\\
\Gamma^2_{33}&=-\sin\theta\cos\theta ,  \qquad
\Gamma^3_{13}=-\frac{M-2\bar{r}}{\bar{r}(M+2\bar{r})},  \qquad
\Gamma^3_{23}=\qopname \relax o{ctg}\theta &\\ \end{split} \]
\end{subsection} 
 \end{document}